\documentclass{PoS}
\usepackage{amsmath}    
\usepackage{graphicx}   
\usepackage{multirow}
\title{Surface operator study in SU(2) gauge field theory}

\ShortTitle{Surface operator study}

\author{Valdimir Goy\\
       School of Biomedicine, Far Eastern Federal University \\
        E-mail: \email{vovagoy@gmail.com}}
        
        \author{Kseniia Durman\\
       School of Biomedicine, Far Eastern Federal University \\
        E-mail: \email{vovagoy@gmail.com}}

\author{\speaker{Alexander Molochkov}\thanks{The project was supported by the Far Eastern Federal University}\\
        School of Biomedicine, Far Eastern Federal University\\
        E-mail: \email{molochkov.alexander@gmail.com}}

\abstract{The surface operator in an SU(2) gauge field theory is studied. We analyze Abelian projection of the SU(2) symmetry to the U(1) group calculating the surface parameter. The surface parameter dependence on the surface area and volume is studied in confinement and deconfinement phases. It is shown the spatial and temporal surface operators exhibit nontrivial area dependence in the confinement and deconfinement phases.  It is shown also that there is no volume law for the operators defined on a cubic surface.}

\FullConference{31st International Symposium on Lattice Field Theory - LATTICE 2014\\
		23-28 June, 2014\\
		Columbia University New York, NY}
\begin{document}
\section{Introduction}
The most important probes for the phase states of a four-dimensional gauge
field theory are the Wilson and t'Hooft line operators that are defined on one-dimensional curves in the space-time\cite{Wilson1974}. For example, these line-operators define order parameters for the confinement-deconfiment phase transition of the QCD vacuum~\cite{CreutzBook1983}. However, for more detail understanding of four-dimensional gauge field theory dynamics and vacuum topology we need additional probes expressed by operators defined on the subspaces with higher dimensions.

 Possible candidates are operators that are defined on two-dimensional hyper-surfaces in the four-dimensional space-time~\cite{Gukov2007}. In the present work  surface operator in an SU(2) non-Abelian gauge field theory is studied.  The surface operator dependence on the surface area and volume is studied in confinement and deconfinement phases. It is shown that both the spatial and temporal surface operators exhibit nontrivial area dependence and no volume dependence for a cubic volume.  

{\section{Surface operator on the lattice}
		
In the present work we study the surface operator that is defined as follows:
\begin{equation}
			W =e^{i\kappa\underset{S}{\oint}F_{\mu\nu}\,d\sigma^{\mu\nu}}
		\label{def}\end{equation}
		where $F_{\mu\nu}$ - the gauge field tensor, $d\sigma_{\mu\nu}$ - surface element, $S$ - a closed hyper-surface. 
For an infinitesimal part of the surface $\Delta S_i^{\mu\nu}$ one can connect the corresponding contribution to the surface integral with the gauge field circulation around the infinitesimal area: 
\begin{equation}
F_{\mu\nu}\,\Delta S_i^{\mu\nu}=\underset{\partial \Delta S_i}\oint A_{\mu}ds^{\mu}
\end{equation}
Within the lattice regularization of the field theory the last integral can be identified with a plaquette:
\begin{equation}\label{plaquette}
\kappa\underset{\partial \Delta S_i}\oint A_{\mu}ds^{\mu}=\theta_i
\end{equation}
As a result the surface integral can be connected with the plaquette angle $\theta_i$ as follows:   
\begin{equation}
\kappa\underset{S}{\int}F_{\mu\nu}\,d\sigma^{\mu\nu}=\underset{S}\sum\theta_i
\end{equation}
Thus, we can define the surface operator~(\ref{def}) on the lattice as follows:
		\begin{equation}\label{lDefW}
			W_{p}\left( S\right) =Re\,\underset{\Delta S_i \in S}{\prod} e^{\imath \theta_{i}},
		\end{equation}
		 Within the pure gauge field theory with $SU\left( 2\right)$, the $\theta_{i}$ related with plaquette value of the gauge filed $F_{i}$ as follows:
		\begin{equation}\label{Fp}
			F_{i}=\widehat{1}\, cos\,\theta_{i} + \imath\,{\bf n}\cdot {\bf \sigma}\, sin\,\theta_{i},
		\end{equation}
		where ${\bf n}$ - vector on the unit sphere, ${\bf \sigma}$ - Pauli matrices, $F_{i}$ is a value of the gauge field tensor $F_{\mu\nu}$ on the plaquette $i$. Thus, for $SU(2)$ projected to $U\left( 1\right)$  one can write the following expression for the $\theta_{i}$~\cite{tHooft1981455}:
		\begin{equation}
			\theta_{i} =\arccos\left( \frac{1}{2}Tr\,F_{i}\right).
		\end{equation}
		
	}
It is important to note that the angle $\theta$ is extracted here in the range $[0,\pi)$, while the expression~(\ref{Fp}) is defined in the range  $[0,2\pi)$ or $[-\pi,\pi]$. This projection will produce additional common factor in the exponent that we do not take into account in this study.     
		To calculate the surface operator on the lattice, we select a cube in the 3d space. Then, the phase is calculated on the each plaquette on the surface of the cube and result is obtained as a sum of these phases. The final result is obtained by averaging the parameter calculated in different points of the lattice configuration and on the set of configurations.
		
		We consider cubes with length of edge from $1a$ to $13a$ ($a$ is the lattice scale), which corresponds area surface from $6$ to $1014$ plaquettes. 
	{\section{Results}
		\begin{figure}
			\includegraphics[width=100mm]{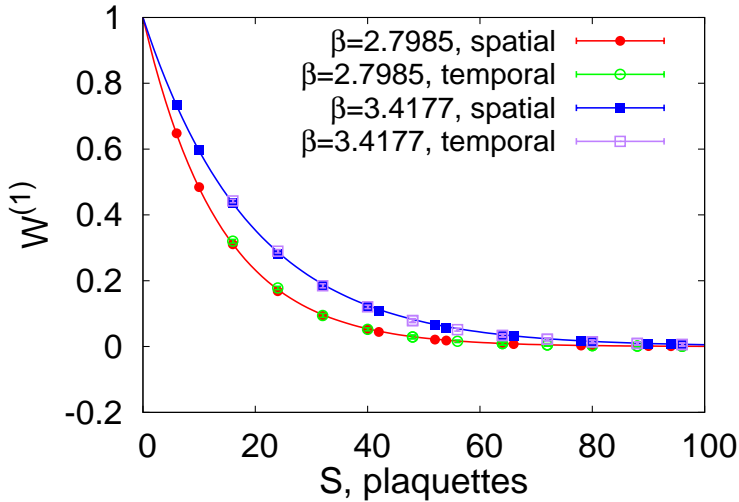}
			\caption{Dependence of the average valuer of the spatial and temporal surface operators from area at the different values of $\beta$ and comparison with the fitting function $e^{-\sigma S}$.}\label{fig:vfit}
		\end{figure}
		All calculation performed on 50 configurations in 1000 points on the each lattice configuration. The results are shown at the figure~\ref{fig:vfit}. To extract area and volume dependence of the surface operator we fit the obtained data by the following expression: 
		\begin{equation}\label{lFitSV}
			W_{p}(S, V)=e^{-\sigma S-\gamma V},
		\end{equation}
		where $\sigma$ is area coefficient, $\gamma$ is volume coefficient, $S$ is the surface area surface, $V$ is volume covered by the surface. The parameters values are obtained with the help of  minuit2 library from ROOT\footnote{See http://root.cern.ch/drupal/} package. 

	\begin{figure}
			\includegraphics[width=100mm]{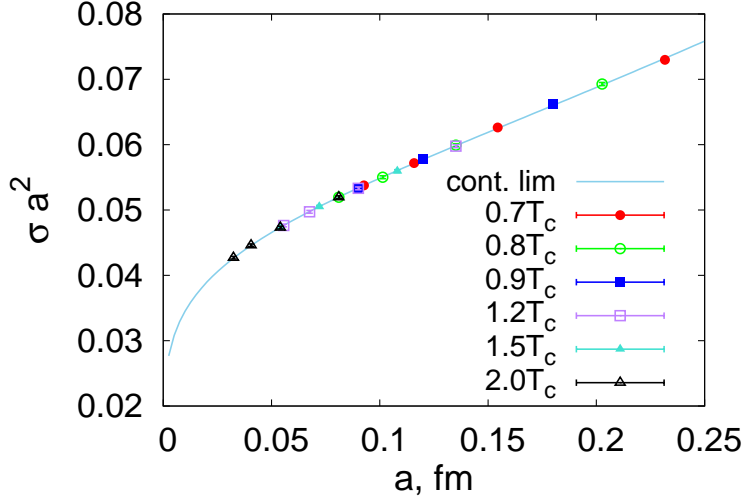}
			\caption{Dependence of the area coefficient $\sigma$ on the lattice spacing. The points are result of the calcuation at the different temperature. Solid line shows fit of the coefficient dependence.}\label{fig:SigmaA}
		\end{figure}

The Figure~\ref{fig:vfit} shows that the surface operator exhibits the area law only ($\gamma=0$). There is no volume law for a cubic surfaces in both phases. Non cubic surfaces show nontrivial volume dependence that needs additional study. 
The Figure~\ref{fig:SigmaA} shows that the surface operator does not depend on the temperature. The area coefficient $\sigma$ was extracted in the following form:

\begin{equation}
\sigma a^2=- c_1+\frac{c_2}{\beta} + c_3 a^2,
\end{equation}
Here, the $c_i$ - constants. Fit of the function (\ref{lFitSV}) gives following values of the constants: $c_1=0.0112013$, $c_2=0.212362$, $c_3=0.158924 fm^{-2}$.
Since there is no temperature dependence of $\sigma$, it has no non-local contributions.
First two terms in this expression are perturbative contributions from the mean plaquette~(\ref{plaquette}), the last term is nonperturbative contribution which has dimension two. 


In conclusion we can say following:
1) the cubic spatial and temporal surface operators exhibit same areal law and no volume law in both confinement and deconfinement phases; 
2) there is no temperature dependence of the cubic surface operators;  3) area coefficient $\sigma$ is expressed via perturbative contributions and nonpertubative contribution of dimension two;
3) volume dependence of the non-cubic surface operators needs more precise study.   
		

	}
	\begin{acknowledgments}
		We are thankful to Valetine Zakharov, Vitaly Bornyakov and Maxim Chernodub for interesting and useful discussions. The work is supported by the Far Eastern Federal University. 
	\end{acknowledgments}

\end{document}